\documentstyle [12pt] {article}

\newcommand{\h}[1]{\mathop{\lambda}\limits_{#1}\ \!\!\!}
\newcommand{\k}[1]{\mathop{h}\limits_{#1}\ \!\!\!}

\newcommand{\edf}{\ {\mathop{=}\limits^{\rm def}}\ }

\newcommand{\al}{\alpha}

\begin{document}
\begin{center}
\bf { SPACE-TIME STRUCTURE AND ELECTROMAGNETISM }
\end{center}
\begin{center}

$\bf{M. I. Wanas ^ {\$,\dag}}$~\& $\bf{Samah. A. Ammar ^ {\#,\dag}}$
\end{center}
\begin{abstract}
Two Lagrangian functions are used to construct geometric field
theories. One of these Lagrangians depends on the curvature of space,
while the other depends on curvature and
torsion. It is shown that the theory constructed from the first
Lagrangian gives rise to pure gravity, while the theory
constructed using the second Lagrangian gives rise to both gravity
and electromagnetism. The two theories are constructed in a
version of absolute parallelism geometry in which both curvature
and torsion are, simultaneously, non-vanishing. One single
geometric object, {\it W-tensor},
reflecting the properties of curvature and torsion, is defined in
this version and is used to construct the second theory. The main
conclusion is that a necessary condition for geometric
representation of electromagnetism is the presence of a
non-vanishing torsion in the geometry used.
\end{abstract}
\section{Introduction}
 In the context of the philosophy of geometerization of physics, any
deviation from  flat space indicates the presence of a type, or
more, of energy which causes this deviation. From the geometric
point of view, the mathematical scheme usually used to detect
whether a space is flat or not is to study the commutation of
tensor derivatives defined in the space. If $A_{\mu}$ is an
arbitrary covariant vector, one can define the tensor derivative
of $A_{\mu}$ as $$A_{\mu|\nu}\edf A_{\mu,\nu}- A_{\al}
\Gamma^{\al}_{.~\mu\nu},\eqno{(1.1)}$$ where the comma (,) denotes
ordinary partial differentiation and $\Gamma^{\al}_{.~\mu\nu}$ is
a linear affine connection defined in the space concerned. In order to
study the commutation of such type of derivatives one should
evaluate the quantity $(A_{\mu|\nu\sigma}-A_{\mu|\sigma\nu})$. If
this quantity vanishes then tensor derivatives commute and the
space is flat, otherwise it is not flat. A general expression for
the above quantity is given by [1]
$$A_{\mu|\nu\sigma}-A_{\mu|\sigma\nu}=A_{\al}B^{\al}_{.~\mu\nu\sigma}-A_{\mu|\al}
~~\Lambda^{\al}_{.~\nu\sigma},\eqno{(1.2)}$$ where
$B^{\al}_{.~\mu\nu\sigma}$ is the \underline{" curvature tensor"} defined by, $$B^{\al}_{.~\mu\nu\sigma}\edf
\Gamma^{\al}_{.~\mu\sigma,\nu}-\Gamma^{\al}_{.~\mu\nu,\sigma}+\Gamma^{\epsilon}_{.`\mu\sigma}
\Gamma^{\al}_{.~\epsilon\nu}-\Gamma^{\epsilon}_{.~\mu\nu}\Gamma^{\al}_{.~\epsilon\sigma}
,\eqno{(1.3)}$$
\underline{~~~~~~~~~~~~~~~~~~~~~~~~~~~~~~~~} \\
$\$$ {\it Astronomy Department, Faculty of Science,}
{\it Cairo University, Giza, Egypt.}

{\it E-mail:wanas@frcu.eun.eg ; mwanas@cu.edu.eg}\\
$\#$ {\it Mathematics}
{\it Department, Faculty of Girls, Ain Shams University, Cairo,}
{\it Egypt.}

{\it E-mail:samahammar@eun.eg}\\
$\dag$ {\it Egyptian Relativity Group (ERG), URL:www.erg.eg.net.}

\newpage
and $\Lambda^{\al}_{.~\nu\sigma}$ is the
\underline{"torsion tensor"} defined by,
$$\Lambda^{\al}_{.~\mu\nu}\edf\Gamma^{\al}_{.~\mu\nu}-\Gamma^{\al}_{.~\nu\mu}
.\eqno{(1.4)}$$

It is clear from (1.2) that both curvature and
torsion can be used as indicators for the deviation of a space
from the flat case. The simultaneous vanishing of both is the
necessary condition for a space to be flat. Consequently, from the
physical point of view, this condition indicates the absence of
any type of energy from the space.

In constructing geometric field theories, using Lagrangian
formalism, one should start with a Lagrangian function. This
Lagrangian is usually formed using a true scalar, i.e. a tensor of
zero order. As it reflects the energetic contents of the space,
the Lagrangian usually constructed from the curvature tensor for
spaces with vanishing torsion (e.g. Riemannian space) (cf. [2]),
or from the torsion tensor in the case of spaces with vanishing
curvature (e.g. Absolute Parallelism space (AP-space)), [3].
What about spaces with, simultaneously, non-vanishing curvature
and torsion? The situation, in this case, is somewhat difficult
since it is not easy to find one single geometric object that
reflects the effects of both curvature and torsion. If such an
object is defined, then the commutation relation (1.2) can be
written in the form, $$
A_{\mu|\nu\sigma}-A_{\mu|\sigma\nu}=A_{\al}W^{\al}_{.~\mu\nu\sigma}.\eqno{(1.5)}$$
In general, the tensor $W^{\al}_{.~\mu\nu\sigma}$ cannot be
defined unless the space is torsion-less. However, there is a
rare case in which such a tensor could be defined in spaces with
simultaneous non-vanishing curvature and torsion. In this case,
this tensor will reflect the effects, of both the curvature and torsion,
on the structure of space. Since this
tensor will represent deviation from flat space and will reduce to
the conventional curvature tensor ( in case of spaces with
vanishing torsion), we are going to call such a tensor the
\underline{" W-tensor"}[4].

It is the aim of the present work to construct and compare field
equations resulting from using the curvature tensor (1.3) and the
W-tensor (1.5) in building  Lagrangian
functions. In Section 2 we are going to review a version of
Absolute Parallelism (AP) geometry in which both curvature and
torsion are simultaneously non-vanishing. Also, we are going to
give the definitions for the curvature and the W-tensors.
In Section 3 the field equations constructed using a Lagrangian built from the
W-tensor are reviewed. A Lagrangian built using the
curvature tensor is used to construct a new set of field equations
in Section 4. The two sets of field equations are compared and
discussed in Section 5.

\section{ Riemann-Cartan Version of AP-Geometry}

The 4-dimensional AP-space is a manifold, each point of which is
labelled by 4-independent variables $x^{\mu}(\mu=0,1,2,3)$. At
each point we define 4-independent contravariant vector $\h{i}^\mu
(\mu =0,1,2,3$ stands for the coordinate components) and
$i(=0,1,2,3$ stands for the vector numbers) \footnote{In the
present work, the authors are going to use the notations in which
the vector number is always written in a lower position (neither
covariant nor contravariant), consequently, such indices are not
lowered or raised ( for more details see [5])}. We use Latin
indices for vector number and Greek indices for coordinate
components. Assuming that $\|\h{i}^{\mu}\|\neq 0$, then we can
define the tetrad covariant vectors $\h{i}_\mu $ as the normalized
cofactors of $\h{i}^\mu $ in the determinant $\|\h{i}^{\mu}\|$,
such that $$ \h{i}^\mu \h{i}_\nu = \delta^{\mu}_{\nu}.
\eqno{(2.1)}
 $$
 Using these vectors, one can define the second order symmetric
tensor,
 $$g_{\mu\nu}\edf \h{i}_{\mu} \h{i}_{\nu}.\eqno{(2.2)}$$
This symmetric tensor can play the role of the metric of
Riemannian space associated with the AP-space. \\

\underline{\bf Connections, Curvatures and Torsion:}

The AP-space admits a non-symmetric linear connection
~~$\Gamma^{\al}_{~~\mu\nu}$, which is a consequence of the
AP-condition [3], i.e. $$\h{i}_{\stackrel{\mu}{+}|~
\nu}\edf\h{i}_{\mu,\nu}-\Gamma^{\al}_{.~\mu\nu}\h{i}_{\al}=0,
\eqno{(2.3)} $$ where the stroke denotes tensor differentiation.
Equation (2.3) can be solved to give,
$$\Gamma^{\al}_{.~\mu\nu}=\h{i}^{\al}\h{i}_{\mu,\nu}.\eqno{(2.4)}$$
Since $\Gamma^{\al}_{.~\mu\nu}$ is non-symmetric, then one can
define its dual connection as
$$\tilde{\Gamma}^{\al}_{.~\mu\nu}\edf\Gamma^{\al}_{.~\nu\mu}.\eqno{(2.5)}$$
The symmetric part of (2.4) is also a connection defined by $$\Gamma^{\al}_{.~(\mu\nu)}\edf\frac{1}{2}(\Gamma^{\al}_{.~\mu\nu}
+\Gamma^{\al}_{.~\nu\mu}).\eqno{(2.6)}$$
Also, Christoffel symbol $\{^{\al}_{\mu\nu}\}$ could be defined
using the metric tensor (2.2). So, in the AP-geometry one could
define four linear connections, at least: the non-symmetric connection
(2.4), its dual (2.5), its symmetric part and Christoffel symbol.

The curvature tensors, corresponding to the above
connections, respectively, are [6]
$$M^{\al}_{.~\mu\nu\sigma}\edf\Gamma^{\al}_{.~\mu\sigma,\nu}-\Gamma^{\al}_{.~\mu\nu,\sigma}
+\Gamma^{\epsilon}_{.~\mu\sigma}\Gamma^{\al}_{.~\epsilon\nu}-\Gamma^{\epsilon}_{.~\mu\nu}
\Gamma^{\al}_{.~\epsilon\sigma},\eqno{(2.7)}$$ $$
\tilde{M}^{\al}_{.~\mu\nu\sigma}\edf
\tilde{\Gamma}^{\al}_{.~\mu\sigma,\nu}-
\tilde{\Gamma}^{\al}_{.~\mu\nu,\sigma}+
\tilde{\Gamma}^{\epsilon}_{.~\mu\sigma}\tilde{\Gamma}^{\al}_{.~\epsilon\nu}-\tilde{\Gamma}^{\epsilon}_{.~\mu\nu}
\tilde{\Gamma}^{\al}_{.~\epsilon\sigma},\eqno{(2.8)}$$
$$\bar{M}^{\al}_{.~\mu\nu\sigma}\edf\Gamma^{\al}_{.~(\mu\sigma),\nu}-\Gamma^{\al}_{.~(\mu\nu),\sigma}
+\Gamma^{\epsilon}_{.~(\mu\sigma)}\Gamma^{\al}_{.~(\epsilon\nu)}-\Gamma^{\epsilon}_{.~(\mu\nu)}\Gamma^{\al}_{.~(\epsilon\sigma)},\eqno{(2.9)}$$
$$R^{\al}_{.~\mu\nu\sigma}\edf\{^{\al}_{\mu\sigma}\}_{,\nu}-\{^{\al}_{\mu\nu}\}_{,\sigma}
+\{^{\epsilon}_{\mu\sigma}\}\{^{\al}_{\epsilon\nu}\}-\{^{\epsilon}_{\mu\nu}\}
 \{^{\al}_{\epsilon\sigma}\}.\eqno{(2.10)}$$ From the AP-condition
(2.3) the curvature tensor given by (2.7) vanishes identically,
while those given by (2.8),(2.9),(2.10) do not vanish.

Using the non-symmetric connection one can define a third order
skew tensor, $$\Lambda^{\alpha}_{.~\mu \nu} \edf
\Gamma^{\alpha}_{.~\mu \nu} - \Gamma^{\alpha}_{.~\nu \mu}=
\tilde{\Gamma}^{\al}_{.~\nu\mu}-\tilde{\Gamma}^{\al}_{.~\mu\nu}
=-\Lambda^{\al}_{.~\nu\mu}.\eqno{(2.11)}$$ This tensor is the
torsion of AP-space. We can define another third order
tensor as,
$$\gamma^{\al}_{.~\mu\nu}\edf\h{i}^{\al}\h{i}_{\mu;\nu},\eqno{(2.12)}$$ where (;) is used to denote covariant differentiation using 
Christoffel symbol. This tensor is called the contortion of the space, using which, it is
shown that (cf. [7])
$$\gamma^{\al}_{.~\mu\nu}=\Gamma^{\alpha}_{.~\mu\nu}-\{^{\alpha}_{\mu\nu}\}.
\eqno{(2.13)}$$ The tensor $\gamma_{\mu\nu\al}$ is skew-symmetric in
its first two indices. A basic vector could be obtained by
contraction, using any one of the above third order tensors, $$
C_{\mu} \edf \Lambda^{\alpha}_{.~\mu \alpha }= \gamma^{\alpha}_{.~
\mu \alpha}. \eqno{(2.14)}$$ Using the contortion, one can define
the symmetric third order tensor as,
$$\Delta^{\al}_{.~\mu\nu}\edf\gamma^{\al}_{.~\mu\nu}+\gamma^{\al}_{.~\nu\mu}.
\eqno{(2.15)}$$ The symmetric and skew-symmetric parts of the
tensor $\gamma^{\al}_{.~\mu\nu}$ are respectively,
$$\gamma^{\al}_{.~[\mu\nu]}=\Gamma^{\al}_{.~[\mu\nu]}=\frac{1}{2}\Lambda^{\al}_{.~\mu\nu},\eqno{(2.16)}$$
$$\gamma^{\al}_{.~(\mu\nu)}=\frac{1}{2}\Delta^{\al}_{.~\mu\nu},\eqno{(2.17)}$$
where the brackets $[~]$ are used for anti-symmetrization and the
parenthesis $(~)$ are used for symmetrization, of tensors, with
respect to the enclosed indices.

Now we have a version of AP-space with, simultaneously,
non-vanishing curvature (2.8) and torsion (2.11) corresponding to the dual connection (2.5). In this space we
can define a W-tensor, of the type given
by (1.5). Since the L.H.S. of (1.5) contains tensor derivatives
using different connections, so we are going to define these
derivatives in the present version of AP-geometry. \\

\underline{\bf Tensor Derivatives}

Using the connections mentioned above, we can define the
following derivatives [6],
$$A^\mu_{+|~ \nu}\edf
A^{\mu}_{~,\nu}+\Gamma^{\mu}_{~\al\nu}A^{\al},\eqno{(2.18)}$$
$$A^\mu_{-|~ \nu}\edf
A^{\mu}_{~,\nu}+\tilde{\Gamma}^{\mu}_{~\al\nu}A^{\al},\eqno{(2.19)}$$
$$A^{\mu}_{0|~\nu}\edf A^{\mu}_{~,\nu}+\Gamma^{\mu}_{~(\al\nu)}A^{\al},\eqno{(2.20)}$$
$$A^{\mu}_{~;\nu}\edf
A^{\mu}_{~,\nu}+\{^{\mu}_{\al\nu}\}A^{\al}.\eqno{(2.21)}$$ As
mentioned in section 1, the definition of a W-tensor, of the type given by (1.5), in spaces with
curvature and torsion, is difficult. The only way, to overcome
this difficulty, is to replace the arbitrary vector $A_{\mu}$ by
the tetrad vectors $\h{i}_{\mu}$ in (1.5). The resulting
definitions can be written as,
$$\h{i}_{\stackrel{\mu}{+}|\nu\sigma}-\h{i}_{\stackrel{\mu}{+}|\sigma\nu}=\h{i}_{\al}
M^{\al}_{.~\mu\nu\sigma},\eqno{(2.22)}$$
$$\h{i}_{\stackrel{\mu}{-}|\nu\sigma}-\h{i}_{\stackrel{\mu}{-}|\sigma\nu}=\h{i}_{\al}
{W}^{\al}_{.~\mu\nu\sigma},\eqno{(2.23)}$$
$$\h{i}_{\stackrel{\mu}{0} | \nu\sigma}-\h{i}_{\stackrel{\mu}{0} | \sigma\nu}=\h{i}_{\al}
\bar{M}^{\al}_{.~\mu\nu\sigma},\eqno{(2.24)}$$
$$\h{i}_{\mu;\nu\sigma}-\h{i}_{\mu;\sigma\nu}=\h{i}_{\al}R^{\al}_{.~\mu\nu\sigma}.\eqno{(2.25)}$$
The first (2.22),  coincides with
the curvature (2.7) and is an identically vanishing tensor
because of (2.3). The tensor given by (2.25) is identical with
Riemannian-Christoffel curvature tensor (2.10) since the connection
used to evaluate the L.H.S. of (2.25) is Christoffel symbol. The symmetric connection (2.6) gives the non-vanishing
W-tensor (2.24) which is identical to the curvature (2.9). So,
we are left with one W-tensor,
${W}^{\al}_{.~\mu\nu\sigma}$, which differs from the
curvature tensors (2.7) - (2.10).

Now multiplying both sides of (2.23) by $\h{i}^{\beta}$, using (2.1), we get [5] $$
{W}^{\al}_{.~\mu\nu\sigma}\edf\h{i}^{\al}(\h{i}_{\stackrel{\mu}{-}|\nu\sigma}
-\h{i}_{\stackrel{\mu}{-}|\sigma\nu}),\eqno{(2.26)}$$  which gives
an explicit definition  W-tensor
defined in the present version of AP-geometry. Now, this version
is characterized by the connection (2.5), the torsion (2.11), the
curvature tensor (2.8) and the W-tensor (2.26). All these geometric objects are , in
general, simultaneously non-vanishing. Such type of geometry, with
simultaneously non-vanishing curvature and torsion, is known as
Riemann-Cartan geometry.

The following table is extracted from [7] and contains second
order tensors that are used in most applications.
\newpage
\begin{center}
 Table 1: Second Order World Tensors [7]      \\
\vspace{0.5cm}
\begin{tabular}{|c|c|} \hline
 & \\
Skew-Symmetric Tensors                &  Symmetric Tensors   \\
 & \\ \hline
 & \\
${\xi}_{\mu \nu} \edf \gamma^{~ ~ \alpha}_{\mu \nu .
|{\stackrel{\alpha}{+}}} $ &
\\

${\zeta}_{\mu\nu} \edf C_{\alpha}~{\gamma^{~~ \alpha}_{\mu \nu .}
} $ &
\\
 & \\ \hline
 & \\
${\eta}_{\mu \nu} \edf C_{\alpha}~{\Lambda^{\alpha}_{.\mu \nu} } $
& ${\phi}_{\mu \nu} \edf C_{\alpha}~\Delta^{\alpha}_{.\mu \nu} $
\\

${\chi}_{\mu \nu} \edf \Lambda^{\alpha}_{. \mu
\nu|{\stackrel{\alpha}{+}} }$ & ${\psi}_{\mu \nu} \edf
\Delta^{\alpha}_{. \mu \nu|{\stackrel{\alpha}{+}}} $
\\

${\varepsilon}_{\mu \nu} \edf C_{\mu | {\stackrel{\nu}{+}}} -
C_{\nu | {\stackrel{\mu}{+}}}$ & ${\theta}_{\mu \nu} \edf C_{\mu |
{\stackrel{\nu}{+}}} + C_{\nu | {\stackrel{\mu}{+}}}  $
\\

${\kappa}_{\mu \nu} \edf \gamma^{\alpha}_{. \mu
\epsilon}\gamma^{\epsilon}_{. \alpha \nu} - \gamma^{\alpha}_{. \nu
\epsilon}\gamma^{\epsilon}_{. \alpha \mu}$   & ${\varpi}_{\mu \nu}
\edf  \gamma^{\alpha}_{. \mu \epsilon}\gamma^{\epsilon}_{. \alpha
\nu} + \gamma^{\alpha}_{. \nu \epsilon}\gamma^{\epsilon}_{. \alpha
\mu}$ \\
 & \\ \hline
 & \\
                  &  ${\omega}_{\mu \nu} \edf \gamma^{\epsilon}_{. \mu \alpha}\gamma^{\alpha}_{. \nu \epsilon}$   \\

                                      &  ${\sigma}_{\mu \nu} \edf \gamma^{\epsilon}_{. \alpha \mu} \gamma^{\alpha}_{. \epsilon \nu}$   \\

                                      &  ${\alpha}_{\mu \nu} \edf C_{\mu}C_{\nu}$   \\

                                      &  $R_{\mu \nu} \edf \frac{1}{2}(\psi_{\mu \nu} - \phi_{\mu \nu} - \theta_{\mu \nu}) + \omega_{\mu \nu}$          \\
 & \\ \hline
\end{tabular}
\end{center}
where $\Lambda^{\al}_{.~\mu \nu |\stackrel{\sigma}{+}} \equiv
\Lambda^{\stackrel{\al}{+}}_{.~\stackrel{\mu}{+}{\stackrel{\nu}{+}}
| \sigma}$. It can be easily shown that there exist an identity
between skew-tensors, of this table, which can be written in the form [7]
$$\eta_{\mu\nu}+\varepsilon_{\mu\nu}-\chi_{\mu\nu}\equiv
0.\eqno{(2.27)}$$ We see from the above table that the torsion
tensor plays an important role in the structure of AP-space in
which all tensors in Table 1 vanish when the torsion tensor
vanishes.

In what follows, we are going to examine and compare the
consequences of constructing field theories, in the AP-geometry,
using the curvature (2.8) and the W-tensor (2.26) to form the corresponding Lagrangian functions.
In 1977 Mikhail and Wanas [8] have constructed the Generalized
Field Theory GFT using the W-tensor (2.26). In
the next section we are going to review briefly this theory, for
the sake of the completeness.
\section{The Use of The W-Tensor}
In constructing GFT, Mikhail and Wanas [8] have used the
W-tensor, in the AP-geometry, (2.23)
which can be written as,
$$\h{i}^{\stackrel{\mu}{-}}_{~|\nu\sigma}-\h{i}^{\stackrel{\mu}{-}}_{~|\sigma\nu}\edf\h{i}_{\al}
W^{\al\mu}_{.~.~\nu\sigma}.\eqno{(3.1)}$$ As stated , the
arbitrary vector $A_{\mu}$ of (1.5) is replaced by the vectors
$\h{i}^{\mu}$ in (3.1).  It is impossible to define The W-tensor unless we make this replacement.  Now,
contracting the tensor $W^{\al\mu}_{.~.~\nu\sigma}$ twice we get
the scalar curvature $W$, using which the Lagrangian scalar
density can be written as,
$$\pounds\edf \lambda^{*}W,\eqno{(3.2)}$$ where
$\lambda^{*}=\|\h{i}_{\mu}\|$ , $L\edf g^{\mu\al}W_{\mu\al}$ and
$$W_{\mu\al}\edf\Lambda^{\epsilon}_{.~\delta\mu}\Lambda^{\delta}_{.~\epsilon\al}
-C_{\mu}C_{\al}.\eqno{(3.3)}$$ It is clear that the tensor
$W_{\mu\al}$ is a symmetric tensor. The theory corresponding to
the scalar density (3.2) is called the GFT which has been
derived by using the variational method of "Dolan and McCrea "
[9]. The field equations of the GFT can be re-derived by
using an action principle method [10]. It is shown that the two
methods give the same set of field equations, which can be written
as, \setcounter{equation}{3}
 \begin{eqnarray}
 E_{\mu\nu}&\edf &g_{\mu\nu}W- 2W_{\mu\nu}-
2g_{\mu\nu}C^{\gamma}_{-~|\gamma}- 2C_{\mu}C_{\nu}\nonumber\\ & &
-2g_{\mu\al}C^{\epsilon}\Lambda^{\al}_{.~\epsilon\nu}
+2C_{\stackrel{\nu}{+}~|\mu}-2g^{\gamma\alpha}\Lambda_{\stackrel{\mu}{+}{\stackrel{\nu}{+}}
{\stackrel{\al}{+}}~|\gamma}=0.\label{(3.4)}
\end{eqnarray} \\

\underline{\bf The symmetric part of $E_{\mu\nu}$:}

The symmetric part of $E_{\mu\nu}$ is defined by,
$$E_{(\mu\nu)}\edf\frac{1}{2}(E_{\mu\nu}+E_{\nu\mu}).\eqno{(3.5)}$$
Using the tensors of Table (1), one can evaluate the definition
(3.5) and gets
$$E_{(\mu\nu)}=g_{\mu\nu}(\sigma-\varpi)+g_{\mu\nu}R-2R_{\mu\nu}-2\sigma_{\mu\nu}
+2\varpi_{\mu\nu}.\eqno{(3.6)}$$ Hence one can write the symmetric
part of the field equations (3.4) in the form $E_{(\mu\nu)}=0$,
i.e.
$$R_{\mu\nu}-\frac{1}{2}g_{\mu\nu}R=\frac{1}{2}g_{\mu\nu}(\sigma-\varpi)
+\varpi_{\mu\nu}-\sigma_{\mu\nu},\eqno{(3.7)}$$ which can be
written in the more compact form,
$$R_{\mu\nu}-\frac{1}{2}g_{\mu\nu}R=T_{\mu\nu},\eqno{(3.8)}$$
where,
 $$T_{\mu\nu}\edf
g_{\mu\nu}\Lambda+\varpi_{\mu\nu}-\sigma_{\mu\nu},\eqno{(3.9)}$$
$$\Lambda\edf\frac{1}{2}(\sigma-\varpi),\eqno{(3.10)}$$
$$ \sigma\edf g^{\mu\nu}\sigma_{\mu\nu},$$ and,
$$\varpi\edf g^{\mu\nu}\varpi_{\mu\nu}.$$
Consequently, by evaluating the vectorial divergence of both sides
of (3.8), it can be easily shown that $$T^{\mu\nu}_{~~;\nu}=0,$$
which gives the conservation of the physical quantities
represented by the tensor $T^{\mu\nu}$. The tensor $T^{\mu\nu}$
can be used to represent the distribution of matter and energy,
i.e. a material-energy tensor. It is to be considered that this
tensor is a geometric object, defined in terms of the building blocks of the AP-structure
used, and not a phenomenological one, in contrast to the case of GR. In the case of the weak field limit,
it has been shown [11] that the symmetric part of the field
equations gives rise to Newtonian gravity. \\

\underline{\bf The Skew Part of $E_{\mu\nu}$:}

The skew part of the field equations (3.4) can be written as
$$E_{[\mu\nu]}\edf\frac{1}{2}(E_{\mu\nu}-E_{\nu\mu})=0.$$ Using
the skew tensors of Table (1), one can write the above equation in
the form,$$F_{\mu\nu}=C_{\mu,\nu}-C_{\nu,\mu},\eqno{(3.11)}$$
where, $$F_{\mu\nu}\edf\zeta_{\mu\nu}+\eta_{\mu\nu}-\xi_{\mu\nu}
.\eqno{(3.12)}$$ Consequently, $F_{\mu\nu}$ will satisfy the
relations,
$$F_{\mu\nu;\sigma}+F_{\sigma\mu;\nu}+F_{\nu\sigma;\mu}=F_{\mu\nu,\sigma}
+F_{\sigma\mu,\nu}+F_{\nu\sigma,\mu}=0.\eqno{(3.13)}$$ It is clear
that equations (3.11) and (3.13) represent a generalization of
Maxwell's field equations, with the skew-symmetric tensor
$F_{\mu\nu}$ representing the electromagnetic field strength, and
the vector $C_{\mu}$ representing the generalized electromagnetic
potential. In the case of the weak field limits, it has been shown
[11] that the skew part of the field equations (3.4) gives rise to
Maxwell's equation .

From the study of the symmetric and the skew parts of the field
equations (3.4) and from applications of the theory [12],
[13], [14], [15], [16], it is clear that the GFT represents a pure geometric theory,
unifying gravity and electromagnetism in the framework of the
geometerization philosophy.

\section{ The Use of The Curvature Tensor }

The curvature tensor (2.8), defined in the context of the structure mentions above, can be written in the
form
$$\tilde{M}^{\beta}_{.~\mu\nu\sigma}\edf\Gamma^{\beta}_{.~\sigma\mu,\nu}-\Gamma^{\beta}
_{.~\nu\mu,\sigma}+\Gamma^{\epsilon}_{.~\sigma\mu}\Gamma^{\beta}_{.~\nu
\epsilon}-\Gamma^{\epsilon}_{.~\nu\mu}\Gamma^{\beta}_{.~\sigma
\epsilon},$$ which is, in general, non-vanishing tensor.
Contracting the above tensor by setting $\beta=\sigma$ we get,
$$\tilde{M}_{\mu\nu}=\Gamma^{\beta}_{.~\beta\mu,\nu}-\Gamma^{\beta}_{.~\nu\mu,\beta}
+\Gamma^{\epsilon}_{.~\beta\mu}\Gamma^{\beta}_{.~\nu
\epsilon}-\Gamma^{\epsilon}_{.~\nu\mu}\Gamma^{\beta}_{.~\beta
\epsilon}.\eqno{(4.1)}$$ Substituting from (2.13) and making some
rearrangements, we can write $$
 \tilde{M}_{\mu\nu}=R_{\mu\nu}-\gamma^{\beta}_{.~\nu\mu;\beta}+\gamma^{\epsilon}_{.~\beta\mu}
 \gamma^{\beta}_{.~\nu \epsilon},\eqno{(4.2)}$$ where $R_{\mu\nu}$ is
 Ricci tensor defined using Christoffel symbol.
The scalar Lagrangian function corresponding to (4.2), can be
defined as $$ \tilde{M}\edf g^{\mu\nu}\tilde{M}_{\mu\nu}.
\eqno{(4.3)}$$ Substituting from definition (4.2) into definition
(4.3) we get, $$ \tilde{M}\edf
g^{\mu\nu}R_{\mu\nu}+C^{\beta}_{.~;\beta}+\frac{1}{2}g^{\mu\nu}[\gamma^{\epsilon}_{.~\beta\mu}
\gamma^{\beta}_{.~\nu
\epsilon}+\gamma^{\epsilon}_{.~\beta\nu}\gamma^{\beta}_{.~\mu
\epsilon}],\eqno{(4.4)}$$ which can be written, using (2.16) and
(2.17), as
$$\tilde{M}=g^{\mu\nu}R_{\mu\nu}+C^{\beta}_{.~;\beta}+\frac{1}{4}g^{\mu\nu}[\Delta^{\epsilon}_{.~\beta\mu}
 \Delta^{\beta}_{.~\nu
 \epsilon}+\Lambda^{\epsilon}_{.~\beta\mu}\Lambda^{\beta}_{.~\nu
\epsilon}].\eqno{(4.5)}$$ The Lagrangian scalar density
corresponding to (4.1) can be defined as, $$\pounds=\lambda^{*}
\tilde{M}.$$ Substituting from (4.5) into the above definition we
get,
$$\pounds=\lambda^{*}g^{\mu\nu}R_{\mu\nu}+\frac{1}{4}\lambda^{*}g^{\mu\nu}
[\Delta^{\epsilon}_{.~\beta\mu}\Delta^{\beta}_{.~\nu
\epsilon}+\Lambda^{\epsilon}_{.~\beta\mu}\Lambda^{\beta}_{.~\nu
\epsilon}]+(\lambda^{*}C^{\beta})_{,\beta}. \eqno{(4.6)}$$ Since
the last term, $(\lambda^{*}C^{\beta})_{,\beta}$, gives no
contribution to the variation, the Lagrangian density (4.6) can be
written in the form
$$\pounds=\lambda^{*}g^{\mu\nu}[R_{\mu\nu}+\frac{1}{4}\Delta^{\epsilon}_{.~\beta\mu}\Delta^{\beta}_{~\nu
\epsilon}+\frac{1}{4}\Lambda^{\epsilon}_{.~\beta\mu}\Lambda^{\beta}_{.~\nu
\epsilon}]=\lambda^{*}\tilde{M} . \eqno{(4.7)}$$ From which it is
clear that
$$\tilde{M}_{(\mu\nu)}\edf R_{\mu\nu}+N_{\mu\nu},\eqno{(4.8)}$$
where, $$N_{\mu\nu}\edf
\frac{1}{4}\Delta^{\epsilon}_{.~\al\mu}\Delta^{\al}_{.~\nu
\epsilon}+\frac{1}{4}\Lambda^{\epsilon}_{.~\al\mu}\Lambda^{\al}_{.~\nu
\epsilon}.\eqno{(4.9)}$$
 To derive the field equations
corresponding to the Lagrangian density function (4.7) we are
going to use the Dolan-McCrea [9] variational method
\footnote{Since Dolan-McCrea method is not published, so the
reader may refer to [8] for more details about its use in the
AP-geometry.}. \\

\underline{\bf Variational Derivatives}

In this method, the Lagrangian function is assumed to be a
function of $\h{i}_{\mu}$ and its first and second derivatives,
i.e.
$$\pounds\equiv
\pounds(\h{i}_{\mu},\h{i}_{\mu,\nu},\h{i}_{\mu,\nu\sigma}).\eqno{(4.10)}$$
Consider now the new function,
$$\pounds_{\eta}\equiv\pounds(\h{i}_{\mu}+\eta \k{i}_{\mu},
\h{i}_{\mu,\nu}+\eta\k{i}_{\mu,\nu}, \h{i}_{\mu,\nu\sigma}+
\eta\k{i}_{\mu,\nu\sigma}),\eqno{(4.11)}$$ where $\eta$ is a small
parameter and $\k{i}_{\mu}$ is a vector field, which will be defined later, in the
AP-space. Then one can write, $$I_{\eta}-I =\int_{\Omega}(
\pounds_{\eta}- {\pounds})dx_{(4)}, \eqno{(4.12)}$$ where
$$dx_{(4)}\edf dx^{0}dx^{1}dx^{2}dx^{3}.$$ and $\Omega$ is a region in the 4-space.
Substituting from (4.10)
and (4.11) into (4.12) and integrating by parts one gets,
$$I_{\eta}-I =\eta\int_{\Omega}(\frac{\delta
L}{\delta\h{i}_{\mu}})\k{i}_{\mu}\lambda^{*}dx_{(4)}+O(\eta^{2}),\eqno{(4.13)}$$
where $$\lambda^{*}\frac{\delta
 L}{\delta\h{i}_{\mu}}\edf\frac{\partial\pounds}{\partial\h{i}_{\mu}}
 -\frac{\partial}{\partial
x^{\nu}}\frac{\partial\pounds}{\partial\h{i}_{\mu,\nu}}+
\frac{\partial^{2}}{\partial
x^{\sigma}x^{\nu}}\frac{\partial\pounds}{\partial\h{i}_{\mu,\nu\sigma}}.
\eqno{(4.14)}$$
 Using (4.14), we can define the new
object,$$S^{\mu}_{.~\nu}\edf\frac{\delta
L}{\delta\h{i}_{\mu}}\h{i}_{\nu}=\frac{1}{\lambda^{*}}\frac{\delta
\pounds}{\delta\h{i}_{\mu}}\h{i}_{\nu},\eqno{(3.15)}$$ where
$\pounds=\lambda^{*} L$. It has been shown [8] that this new object
is a tensor of the character indicated by its indices . \\

\underline{\bf An Integral Identity}

We are going to use the infinitesimal transformation
$$\bar{x^{\mu}}=x^{\mu}+\eta\xi^{\mu},\eqno{(4.16)}$$ where $\eta$
is the small parameter, mentioned above, whose square and
higher orders can be neglected. Now, we calculate $\bar{\h{i}_{\mu}}$ in terms of $\h{i}_{\mu}$ using two
methods. The first method, is by using the tensor properties, we have
$$\bar{\h{i}_{\mu}}(\bar{x})=\frac{\partial x^{\nu}}{\partial
\bar{x}^{\mu}}\h{i}_{\nu}(x).$$ Substituting from (4.16) into
the above equation we get,
$$\bar{\h{i}_{\mu}}(\bar{x})=\h{i}_{\mu}(x)+\eta\h{i}_{\nu}z^{\nu}_{~,\mu}+O(\eta^{2}).\eqno{(i)}$$
\\ The second is by using Taylor's expansion, we can write
$$\bar{\h{i}_{\mu}}(\bar{x})=\bar{\h{i}_{\mu}}(x-\eta
z)=\bar{\h{i}_{\mu}}(x)-\eta\h{i}_{\mu,a}z^{a}+O(\eta^{2}.)\eqno{(ii)}$$
From $(i)$ and $(ii)$, we can write
$$\bar{\h{i}_{\mu}}(x)-\h{i}_{\mu}(x)=\eta\k{i}_{\mu}+O(\eta^{2}),\eqno{(4.17a)}$$
where,
$$\k{i}_{\mu}\edf\h{i}_{\nu}~z^{\nu}_{~,\mu}+\h{i}_{\mu,\nu}~z^{\nu}=\h{i}_{\nu}
~z^{\nu}_{-|~\mu}.\eqno{(4.17b)}$$

Also, one can treat $\bar{\pounds}(\bar{x})$ in a similar manner, in
the following:
 \\ First, we have   $$\begin{array}{cccc}
\bar{\pounds}(\bar{x})&=&\bar{\pounds}(x-\eta z)   \\
  &=&\bar{\pounds}(x)-\eta\pounds(x)_{,a}z^{a} &+O(\eta^{2}).
\end{array}\eqno{(iii)}$$
 The function $\bar{\pounds}(x)$ stands for
$$\bar{\pounds}(x)\equiv\bar{\pounds}(\bar{\h{i}_{\mu}}(x),\bar{\h{i}_{\mu,\nu}}(x),
\bar{\h{i}_{\mu,\nu\sigma}}(x)).\eqno{(iv)}$$ Secondly, we know
that the function $L$ is a scalar so,
$$\bar{L}(\bar{x})=L(x),$$ thus,
$$\bar{\lambda^{*}}\bar{L}(\bar{x})=\bar{\lambda^{*}}
L(x)=\lambda^{*}L(x)\frac{\partial(x)}{\partial(\bar{x})}.\eqno{(v)}$$
Using equation (4.16), one can write the equation $(v)$ in the
form,
$$\bar{\pounds}(\bar{x})=\pounds(x)+\eta\pounds(x)z^{a}_{~,a}+O(\eta^{2}),\eqno{(vi)}$$
comparing $(iii)$ and $(vi)$ , one can write
$$
\int_{\Omega}(\bar{\pounds}(x)-\pounds(x))dx_{(4)}=\eta\int_{\Omega}(\pounds(x)z^{a})_{,a}dx_{(4)}
+O(\eta^{2}),\eqno{(vii)}$$
 since,
 $$\int_{\Omega}(\pounds(x)z^{a})_{,a}dx_{(4)}=\int_{\Sigma}\pounds(x)n_{a}z^{a}d\Sigma\equiv
 0.$$ Gauss's theorem has been used to convert the volume integral
 to a surface integral in which $n_{a}$ is a unit vector normal to
 hypersurface $\Sigma$. The integral vanishes because $z^{a}$ vanishes
 at all points of $\Sigma$. Then we can write equation $(vii)$ in the form, $$\int_{\Omega}(\bar{\pounds}(x)-\pounds(x))dx_{(4)}=O(\eta^{2})
 .$$  \\
If we substitute for $\bar{\h{i}_{\mu}}(x)$
 from (4.17a) into $(iv)$ using (4.11), we get
$$\bar{\pounds}(x)=\pounds_{\eta}(x)+O(\eta^{2}).\eqno{(viii)}$$
Substituting from $(viii)$ into$(vii)$ we get,
$$\int_{\Omega}(\pounds_{\eta}-\pounds)dx_{(4)}=0(\eta^{2}).\eqno{(4.18)}$$
Comparing the orders of magnitude of different terms on the R.H.S.
of (4.12) and (4.18), we get the identity
$$\int_{\Omega}(\frac{\delta
L}{\delta\h{i}_{\mu}})\k{i}_{\mu}\lambda^{*}dx_{(4)}\equiv
0.\eqno{(4.19)}$$  Substituting from (4.17b) and (4.15) into (4.19)
we get the differential identity,
$$S^{\stackrel{\mu}{-}}_{.~\stackrel{\nu}{-}|~\mu}\equiv
0.\eqno{(4.20)}$$ \\
\underline{\bf Field Equations}

Considering the identity (4.20) as a generalization of contracted Bianchi
 second identity implying conservation, a set of field equations
corresponding to this identity can be taken as,
$$S^{\mu}_{.~\nu}=0.\eqno{(4.21)}$$
We can write the field equations resulting from the Lagrangian
function (4.7) by using (4.15) as,
$$S^{\beta}_{.~\sigma}\edf\frac{\delta \tilde{M}}{\delta
\h{j}_{\beta}}\h{j}_{\sigma},\eqno{(4.22)}$$ where, $$\frac{\delta
\tilde{M}}{\delta\h{j}_{\beta}}\edf\frac{1}{\lambda^{*}}(\frac{\partial\pounds}{\partial\h{j}_{\beta}}
-\frac{\partial}{\partial x^{\gamma}}\frac{\partial
\pounds}{\partial\h{j}_{\beta,\gamma}}+\frac{\partial^{2}}{\partial
x^{\mu}\partial
x^{\gamma}}\frac{\partial\pounds}{\partial\h{j}_{\beta,\mu\gamma}}).\eqno{(4.23)}$$
Now using (4.7), (4.22),(4.23) in (4.21) and performing necessary
calculations, we get after some rearrangements,
\setcounter{equation}{23}
 \begin{eqnarray}
S^{\beta}_{.~\sigma}&\edf &-2G^{\beta}_{.~\sigma}+N
\delta^{\beta}_{\sigma}-2N^{\beta}_{.~\sigma}
+2\gamma^{\gamma\beta}_{.~.~\sigma|\stackrel{\gamma}{+}}+2\gamma^{\epsilon\beta}_{.~.~\al}
\gamma^{\al}_{.~\sigma\epsilon}\nonumber \\ & &
+\gamma^{\al\beta}_{.~.~\gamma}\gamma^{\gamma}_{.~\al\sigma}-\gamma_{\al}^{.~\epsilon\beta}\gamma^{\al}_{.~\sigma
\epsilon}-2C_{\al}\gamma^{\al\beta}_{.~.~\sigma}. \label{(4.3)}
\end{eqnarray}
The above equation represents a new set of field equations
corresponding to the Lagrangian function (4.7). To extract physical meanings from the geometric equation (4.24), one should
study the symmetric and the skew parts of this equation. For this reason, we are going to write (4.24) in the form,
 \setcounter{equation}{24}
 \begin{eqnarray}
 S_{\nu\sigma}&\edf &-2 G_{\nu \sigma}+N g_{\nu\sigma}-2N_{\nu\sigma}
+2\gamma^{\gamma}_{.~\nu\sigma|\stackrel{\gamma}{+}}+2\gamma^{\epsilon}_{.~\nu\mu}
\gamma^{\mu}_{.~\sigma \epsilon}\nonumber \\ & &
+\gamma^{\al}_{.~\nu\gamma}\gamma^{\gamma}_{.~\al\sigma}+\gamma^{\epsilon}_{.~\mu
\nu}\gamma^{\mu}_{.~\sigma
\epsilon}-2C_{\al}\gamma^{\al}_{.~\nu\sigma}=0.\label{(4.39)}
 \end{eqnarray} \\

 \underline{\bf The Symmetric Part of $S_{\nu\sigma}$}

 The symmetric part of $S_{\nu\sigma}$ is defined as usual by,
$$S_{(\nu\sigma)}\edf\frac{1}{2}(S_{\nu\sigma}+S_{\sigma
\nu}).$$Substituting from (4.25) into the above definition and
using the symmetric tensors of Table (1), we can write
$$S_{(\nu\sigma)}=-2 G_{\nu\sigma}-g_{\nu\sigma}\omega
+\psi_{\nu\sigma}+2~\omega_{\nu\sigma}-\phi_{\nu\sigma} = 0,
\eqno{(4.26)}$$ which can be written in the more convenient form,
$$
R_{\nu\sigma}-\frac{1}{2}g_{\nu\sigma}R=(\frac{1}{2})(-g_{\nu\sigma}\omega+\psi_{\nu\sigma}
+2~\omega_{\nu\sigma}-\phi_{\nu\sigma}).\eqno{(4.27)}$$  If we
define the tensor
$$\tilde{T}_{\nu\sigma}\edf(\frac{1}{2})(-g_{\nu\sigma}\omega+\psi_{\nu\sigma}
+2~\omega_{\nu\sigma}-\phi_{\nu\sigma}),\eqno{(4.28)}$$ then we
can write (4.27) in the form
$$R_{\nu\sigma}-\frac{1}{2}g_{\nu\sigma}R=\tilde{T}_{\nu\sigma},$$
from which it is clear that,
 $$ \tilde{T}^{\nu\sigma}_{~~;\sigma}=0.\eqno{(4.29)}$$ This
implies conservation (since the vectorial divergence of the left
hand side of (4.27) vanishes identically). Consequently, $\tilde{T}^{\nu\sigma}$ can be taken to represent
the material-energy distribution in the present theory. In the case of weak
field limits, the symmetric part of the field equations will tend
to
$$R^{(1)}_{\nu\sigma}-\frac{1}{2}\delta_{\nu\sigma}R^{(1)}=\psi^{(1)}_{\nu\sigma}\eqno{(4.30)}$$
 the index (1) is used to indicate linearity. This equation gives rise to a gravitational field (compare with the
similar case in GR) within a material distribution, characterized by
$\psi^{(1)}_{\nu\sigma}$. \\

\underline{\bf The Skew-Symmetric Part of $S_{\nu\sigma}$}

To obtain the skew part of the field equations (4.25) we use, as usual the
definition,
$$S_{[\nu\sigma]}\edf\frac{1}{2}(S_{\nu\sigma}-S_{\sigma\nu}) = 0
.$$ Substituting from (4.25) into the above definition and using
the skew tensors of Table (1) we
get,$$S_{[\nu\sigma]}=\chi_{\nu\sigma}-\eta_{\nu\sigma}= 0.$$ This
can be written as,
$$\eta_{\nu\sigma}-\chi_{\nu\sigma}=0.\eqno{(4.31)}$$ Now using
the identity (2.27), we get $$\epsilon_{\nu\sigma}=0,
\eqno{(4.32)}$$ which can be written in the form,
$$\eta_{\nu\sigma}=C_{\nu,\sigma}-C_{\sigma,\nu} .\eqno{(4.33)}$$
 We can take the tensor $\eta_{\nu\sigma}$ to represent the strength of the
 electromagnetic field
whose generalized potential is $C_{\mu}$, as usually done in the
literature (cf. [8]), then equation (4.33), apparently, indicates
the presence of an electromagnetic field. In the case of the weak
field limits it can be shown that the skew part of the field
equations (4.25) will give rise to the equation,
$$C^{(1)}_{\nu,\sigma}- C^{(1)}_{\sigma, \nu}=0,\eqno{(4.34)}$$
where $C^{(1)}_{\nu}$ denotes the linear part of the vector
$C_{\nu}$. This equation implies the absence of the
electromagnetic field in the weak field limit.

\section{ Discussion And Concluding Remarks}
 In the present work, we have directed the attention to a version
 of AP-geometry that admits spaces with, simultaneously
 non-vanishing,
 curvature and torsion. Such type of spaces is known in the
 literature as Riemann-Cartan spaces. The main characteristic of a version of this type is,
  briefly,
 reviewed in Section 2. The structure of this space depends on the
 dual connection (2.5), using which a curvature (2.8) the
 torsion (2.11) and the W-tensor (2.26) are defined. It is clear that these geometric
 objects are, in general, simultaneous non-vanishing tensors. It is shown that, in this type
 of spaces, the fourth order tensor (2.26), which is different from the
curvature (2.8), the W-tensor has been first defined [5] and used [8] to construct the GFT.
 It has been classified [11] as a pure geometric attempt unifying
 gravity and electromagnetism. Several applications of this theory
 support this classified (cf. [12],[13],[14],[15],[16]). For the sake of comparison,
  this theory is briefly reviewed in Section 3.

 In Section 4 the authors have constructed a new field theory,
 using the dual curvature tensor (2.8) to form the
 Lagrangian function of the theory. It can be easily shown that
 this tensor is a non-vanishing one, since it is defined in terms
 of the dual connection (2.5) of the AP-space. The variational
 method used, to obtain the field equations of this theory, is
 that of Dolan and McCrea. This method depends mainly on
 the comparison of the same order of magnitude of different terms
 of an integral, that has been evaluated  using two different
 expansions. This method is, somewhat, different from the action
 principle method, although the two methods give identical results
 (compare [8] and [10]).

 The number of field equations of the new theory (4.25) are sixteen.
 Ten of which (4.27) are symmetric and rest (4.33) are
 anti-symmetric. Conservation of matter and energy is guaranteed
 from (4.29). The weak field approximation of the antisymmetric
 part of the field equations indicates the absence of,
 conventional,
 electromagnetic fields. So, the theory, at least, can represent  pure gravity and in this
 case the antisymmetric part of the field equations, (4.33), are
 used to fix the extra six degrees of freedom of the tetrad (since the field variables
  are sixteen, the tetrad vectors). As it is shown [11] the generalized electromagnetic
  potential is represented, geometrically, using the basic vector (2.14) which is the trace
   of the torsion. In the
  new theory, derived in Section 4, this potential does not vanish
  even in the weak field limit. But since this theory is a pure
  gravity theory as shown, then this potential can be classified
  among a type of potentials that do not generate electromagnetism
  and cannot be removed by any transformation [17]. This type of
  potentials may give rise to an alternative interpretation of the
  Aharonov-Bohm effect.

  The advantage of using such theories, in applications, is its pure geometric object (4.28)
  that can describe the material-energy distribution. The use of this object may overcome the difficulty
  of imposing an equation of state, from outside the theory. This may throw more light on the gravitational field and the
  physical situation within material distributions.
 \begin{center}
Table 2: Comparison Between GFT and The New Theory \\
 \vspace{0.5cm}
 \begin{tabular}{|c|c|c|}\hline & & \\
   Criteria & GFT & The New Theory \\ & & \\ \hline & & \\
   Geometry & AP & AP \\ & & \\ \hline & & \\
   Lagrangian & $
   g^{\mu\al}(\Lambda^{\beta}_{.~\delta\mu}\Lambda^{\delta}_{.~\beta\al}
   -C_{\al}C_{\mu})$ & $g^{\mu\nu}(R_{\mu\nu}+\frac{1}{4}\Delta^{i}_{.~\al\mu}
   \Delta^{\al}_{.~\nu i}$ \\
   &(3.2) &$+\frac{1}{4}\Lambda^{i}_{.~\al\mu}\Lambda^{\al}_{.~\nu i})~~~~ (4.7)$ \\& & \\ \hline & & \\
   Tensor used & W-tensor (2.26) &  dual curvature (2.8) \\ & & \\ \hline & & \\
   Field equations & non-symmetric & non-symmetric \\ & & \\ \hline & & \\
   Material-energy tensor & geometric & geometric \\ & & \\ \hline & & \\
   Interactions &  gravitational and & gravity with a geometric  \\
    (weak field regime)& electromagnetic & matter-tensor \\ & & \\ \hline & & \\
   Identities & generalized Bianchi & generalized Bianchi \\ & & \\ \hline
 \end{tabular}
 \end{center}

 Table 2 gives a comparison
  between the GFT and the new theory derived in the present work.
  From this Table, it is clear that the use of the curvature tensor in
   constructing the Lagrangian (4.6) leads to a pure geometric gravity theory, in which conventional electromagnetism is absent.
   While the use of the W-tensor (3.1), for
   the same purpose, leads to a theory for gravity and
   electromagnetism. It is to be considered that torsion does not enter
   the Lagrangian (3.2) as an added entity, but this Lagrangian contains curvature
   and torsion, fused as one  entity, in the W-tensor given by (3.1).
    The difference between the two tensors is that
   the curvature one reflects the effect of the curvature,
   only, on the properties of the space; while the W-tensor (3.1) reflects
   the effects of both curvature and torsion on the structure of
   the space. This may lead to the conclusion that torsion and
   electromagnetism are closely related. Further investigations are needed to support
   this conclusion and to explore the type of relation between torsion and electromagnetism.

   To summarize, and as stated in the introduction, we have: On
   one hand, the energy of any system is represented by its
   Lagrangian, which is a scalar density. On the other hand, the
   presence of energy in a space will deviate this space from the
   flat case. We have examined and compared the consequences of constructing
   field theories using two different Lagrangian functions:

   1- The first is built up using the W-tensor
   (2.23), Section 3.

   2- The second is built up using the dual curvature
   (2.8), Section 4.\\
   Note that both depend on the same linear connection, the dual connection (2.5).
   It is shown, in the present work, that the existence of
   electromagnetism is closely related with the first Lagrangian
   (3.2), not with the second one (4.7). Since the first
   Lagrangian implies the effect of both curvature and torsion, on
   the structure of space time; while the second implies the
   effect of the curvature only, one can conclude that: {\bf To unify
   gravity and electromagnetism, in a 4-dimensional space-time, the
   geometry should admit a non-vanishing torsion that, together with
   curvature, is to be used to build the Lagrangian of the theory}. This is
   to guarantee that more types of energy, including the
   electromagnetic energy, are taken into consideration.

\section*{Acknowledgements}

The authors are indebted to members of the Egyptian Relativity Group (ERG) for many discussions.

\section*{References}

 { [1] Eisenhart, L.P. (1927) {\it "Non-Riemannian Geometry"} AMS,
 New York.} \\
 { [2] Adler, R., Bazin, M. and Schiffer, M. (1975) {\it "Introduction
to General

Relativity"}, 2nd.ed., McGraw-Hill, New York.}\\ { [3] Einstein,
A. (1930) Math. Annal. {\bf 102}, 685.}\\
{ [4] Youssef, N.L. and Sid Ahmed, A.M. (2007) Rep. Math. Phy. {\bf 60}, 39.} \\
{ [5] Wanas, M.I. (1975) Ph.D. Thesis, Cairo University. } \\
{ [6] Wanas, M.I. (2001) Stud.Cercet.Stinn.Ser.Mat.Univ.Bacaau {\bf 10}, 297, gr-qc/0209050}\\
{ [7] Mikhail, F.I. (1962) Ain Shams Sci. Bul. {\bf 6}, 87.}
\\ { [8] Mikhail,
F.I. and Wanas, M.I. (1977) Proc. R. Soc. Lond. A. {\bf 356},
471.}\\ { [9] Dolan, P. and McCrea, W.H. (1963) {\it "Personal
Communications"}.} \\ { [10] Mikhail, F.I. and Wanas, M.I. (1998)
gr-qc/9812086. } \\   { [11] Mikhail, F.I.
and Wanas, M.I. (1981) Int. J. Theor. Phys. {\bf 20}, 671.}\\
 { [12] Wanas, M.I. (1981) Nuovo Cimento, {\bf B.66}, 145.} \\ {
 [13] Wanas, M.I. (1985) Int. J. Theor. Phys. {\bf 24}, 639.} \\ {
 [14] Wanas, M.I. (2007) Int.Jou.Geom.M.Mod.Phy. {\bf 3}, V.4, 373} \\ { [15]
 Wanas, M.I. (1989) Astrophys. Space Sci. {\bf 154}, 165.} \\ {
 [16] Mikhail, F.I., Wanas, M.I. and Eid, A.M. (1995) Astrophys. Space
 Sci. {\bf 228}, 221.} \\ { [17] Ageev, A.N. and Chirkov, A.G. (2002)
gr-qc/0205081. }

\end{document}